\documentclass[11pt]{article}
\usepackage{graphicx,amssymb,amsmath}

\begin{document}

\newcommand\eq[1]{Eq.~(\ref{eq:#1})}
\thispagestyle{empty}
\begin{titlepage}
\begin{flushright}
CALT-TH-2016-007
\end{flushright}
\vspace{1.0cm}
\begin{center}

\bigskip
\bigskip
{\LARGE \bf    Global Gauge Symmetries, Risk-Free Portfolios, and the Risk-Free Rate}\\ 
\bigskip
\bigskip

\bigskip
{Martin Gremm } \\
~\\
\noindent
{\it\ignorespaces
Pivot Point Advisors, LLC\\
 5959 West Loop South, Suite 333,  Bellaire, TX 77401} \\
{\tt  gremm@pivotpointadvisors.com }\\
\bigskip

\bigskip
\end{center}
\vspace{1cm}
\begin{abstract}
We define risk-free portfolios using three gauge invariant differential operators that require such portfolios to be insensitive to price changes, to be self-financing, and to produce a zero real return so there are no risk-free profits. This definition identifies the risk-free rate as the return of an infinitely diversified portfolio rather than as an arbitrary external parameter. The risk-free rate  measures the rate of global price rescaling, which is a gauge symmetry of economies. We explore the properties of risk-free rates, rederive  the Black Scholes equation with a new interpretation of the risk-free rate parameter as a that background gauge field, and discuss gauge invariant  discounting of cash flows.

\end{abstract}
\vfill

\end{titlepage}

\section{Introduction}

Risk-free portfolios play a central role in finance. Their  returns are used to discount cash flows. Many option pricing equations are derived by constructing a risk-free portfolio out of the option and the underlying asset, and setting the return of this portfolio equal to the risk-free rate. The risk-free rate is also the natural benchmark for investment returns. 

Despite their importance,  there is  no universal definition of risk-free portfolios.  When discounting cash flows, risk-free usually means free from default risk. In arbitrage arguments it usually means insensitive to small price fluctuations of the portfolio components. These two definitions are incompatible and cannot appear in the same calculation without causing inconsistencies. 

In much of the literature, the risk-free rate appears as an arbitrary external parameter. This is unsatisfactory precisely because the risk-free rate is such a fundamental component of any economy. The properties of economies should determine the risk-free rate. Here we propose a method for constructing a portfolio that yields the risk-free rate out of the assets that make up an economy. 

We define risk-free portfolios to be  insensitive to price fluctuations, produce zero real return, and to be self-financing. The first requirement equates risk with the uncertainty from  random price changes. The second requirement ensures that there are no risk-free profits, as required for markets in equilibrium. The third requirement  allows us to use the return of any risk-free portfolio to discount cash flows.

Default-free bonds are risky because they do not satisfy the first condition, but hedged options portfolios satisfy all requirements by construction. There is also a class of infinitely diversified portfolios that satisfy our definition. To the best of our knowledge, this type of risk-free portfolio as not been discussed in the literature. We show that the return of these portfolios appears naturally where one expects to see the risk-free rate. This identification leads to an interpretation of the risk-free rate as a measure of global price  inflation.

The implementation of the three requirements that define risk-free portfolios must respect  global symmetries of economies in equilibrium. We consider the two most fundamental symmetries that exist in all economies. The first is a global rescaling of prices. If the buying power of each dollar doubles and the cost of all goods also doubles, the economy remains unchanged because the owner of a dollar can still buy the same quantity of goods as before.  The second global symmetry is the freedom to redefine the unit of trade. A simple example is a stock split. It is a change of unit, but it has no economic effect. 

The global price rescaling symmetry was first discussed in \cite{malaney1996}.  Shortly after  \cite{ilinski1997} promoted it to a local symmetry and argued that the resulting model is a description of markets out of equilibrium.   \cite{young1999} provides a map between lattice gauge theories and foreign exchange markets.  \cite{hoogland2001} and \cite{zhou2010} discusses options prices in the context of  the price rescaling symmetry. The evolution of portfolios in the context of local gauge invariance is discussed, e.g., in \cite{farinelli2015}. Reviews of  gauge symmetries in finance can be found in \cite{smolin2009} and \cite{maldacena2014}.

In next section  we discuss the two  global symmetries mentioned above. Section \ref{definition} presents the differential operators that define  risk-free portfolios. Section \ref{rfr} discusses the properties  of asymptotically risk-free portfolios including their relationship to the price rescaling symmetry. The differences between gauge symmetries and numeraire changes are the subject of Section \ref{numerairetransformation}. In Section \ref{options} we show that the Black Scholes equation \cite{black1973} is a special case of our definition of risk-free portfolios. We also discuss a simple generalization to incorporate discounting with approximately  risk-free portfolios. In Section \ref{discount} we derive a gauge invariant  discount factor for cash flows received in the future. We briefly comment on the effect of parameter uncertainty in Section \ref{sensitivity} before offering a few concluding  remarks in the last section.

\section{Gauges Symmetries}
\label{gaugesymmetries}

Our economic universe contains $N$ assets with  prices, $s_i$, and quantities, $q^i$. The prices are functions of time. The quantities are functions of the prices and time. The value of a portfolio is  $\mathbf{s}\cdot\mathbf{q} = s_iq^i$. 
In addition there are  $M$ derivatives on the assets with prices $V_l$ and quantities $Q^l$. A universe containing derivatives and underlying assets has a price vector  $\mathbf{P} = (\mathbf{s}, \mathbf{V})$ and a corresponding quantity vector  $\mathbf{K} = (\mathbf{q}, \mathbf{Q})$. The value of a portfolio containing both derivatives and underlying assets is given by $\Pi =\mathbf{P}\cdot\mathbf{K} = P_\alpha K^\alpha$. Summation over raised and lowered indices is implied.


Economies are invariant under a possibly time-dependent global rescaling of  prices,
\begin{equation}
\label{scaletransformation}
s_i \to s_i^\prime = e^{\phi(t)}s_i,
\end{equation}
because this leaves all exchange rates unchanged. The option prices, $V_l(s_i)$, scale the same way. The gauge parameter, $\phi(t)$ is an arbitrary deterministic function of time. A choice of $\phi(t)$ is often called a gauge choice. A quantity or expression that does not depend the choice of $\phi(t)$ is gauge invariant. 

In the literature, this gauge invariance is often equated with the freedom to choose a numeraire, i.e., the freedom to make one price the unit of measurement. As we show in Section \ref{numerairetransformation}, these transformations are related, but not identical. A numeraire transformation changes the stochastic properties of prices, while a gauge transformation leaves them invariant. Unless stated otherwise, the results presented here rely on the more general gauge invariance. 

The second global gauge symmetry is a redefinition of the unit of trade. A familiar example is a stock split. The unit of trade changes, usually by a factor of two, and the price per unit changes to ensure that the value of a position in this stock does not change as the split occurs. We can redefine the unit of trade more generally, for example by agreeing to trade certain linear combination of assets, instead of single assets.  The gauge group for these transformations is $GL(N)$, the group of general linear transformations. A time-dependent $b^i_j \in GL(N)$ acts on quantities and prices as
\begin{equation}
\label{qtransformation}
q^i \to q^{\prime i} =b^i_j q^j \qquad\text{and}\qquad s_i \to s^\prime_i={b^{-1}}^j_i s_j.
\end{equation}
There is a similar transformation on the space of options.

For later use we note how portfolios transform under the gauge symmetries discussed here. A portfolio $\Pi=\mathbf{P}\cdot\mathbf{K}$ is invariant under redefinitions of the unit of trade, Eq.~(\ref{qtransformation}), but it transforms as $\Pi\to\Pi^\prime= e^{\phi(t)} \Pi$ under the rescaling summetry, Eq.~(\ref{scaletransformation}). Correspondingly, the portfolio return transforms as
\begin{equation}
\label{pitransform}
\frac{\dot{\Pi}}{\Pi} \to \frac{\dot{\Pi}}{\Pi} + \dot{\phi}.
\end{equation}

It is useful to categorize observables into those that depend on the gauge choice and those that are gauge-independent. The latter are economically relevant because everyone agrees on their properties regardless of the gauge they use. Gauge-dependent quantities cannot be economically relevant, because each observer is free to chose a different gauge and thereby a different value for those observables. Portfolio returns are gauge-dependent as can bee seen from  Eq.~(\ref{pitransform}). However, the difference of two returns is an example of a gauge-independent quantity that contains economically relevant information.

\section{Defining Risk-Free Portfolios}
\label{definition}

Our definition of risk-free portfolios follows the definition used in arbitrage arguments where risk is generally equated with uncertainty. In our universe, the only uncertainty comes from the random fluctuations of asset prices. Portfolios that are insensitive to (small) changes in prices evolve deterministically and are conventionally called risk-free. We will refer to them as price-insensitive, because  risk-free portfolios must satisfy additional requirements. 

The second component of arbitrage arguments  is that markets in equilibrium offer no opportunities to make a risk-free profit. If such arbitrages existed, traders would take advantage of the them until they disappear, returning markets to equilibrium. Consequently we require that investing  in a risk-free portfolio does not yield an economic benefit. 

Finally, we require that risk-free portfolios must be  self-financing so they can be used to discount asset  flows. We simply buy the risk-free portfolio, hold it for some time without adding to or withdrawing from our investment, and sell it. The proceeds at the final time are the time-translated value of the initial purchase price. 

In summary, we define risk-free portfolios as portfolios that satisfy three requirements: They are insensitive to price changes, holding them yields no economic benefit, and they are self-financing. In this section we translate this definition into a set of differential operators that respect the global symmetries of economies. 

The first requirement is that risk-free portfolios must be insensitive to price changes. This is the case if the quantity vector, $\mathbf{K}$,  satisfies
\begin{equation}
\label{riskfree}
 K^\alpha  \frac{\partial}{\partial s_i}P_\alpha= 0
\end{equation}
at all times. Evaluating this on a universe with $N=M=1$ we find that the portfolio is insensitive to price changes if

\begin{equation}
  q^1 + Q^1 \frac{\partial V_1}{\partial s_1} = 0,
\end{equation}
where  $s_1$ is the price of the single asset,  $q^1$ the corresponding quantity,  $V_1$ the option price, and $Q^1$ the corresponding quantity. This is the standard delta hedging requirement. 

Consider a universe with $N$ assets and  no derivatives. Then the components of the quantity vector  must satisfy 
\begin{equation}
\label{riskfreeport}
q^j\frac{\partial}{\partial s_i} s_j = q^i = 0.
\end{equation}
For finite $N$ the only solution is $\mathbf{q}=0$, but in the $N\to\infty$ limit additional solutions appear. If the prices are independent, any portfolio with $q^i$ that  scale as  $1/N$  is asymptotically insensitive to price changes because the exposure to each asset tends to zero. We show that these infinitely diversified portfolios  return the risk-free rate. 

Prices within a given asset class are often correlated, violating our assumption of independence. However, we can construct risk-free portfolios if an economy with $N\to\infty$ assets contains an infinity of correlated asset sets that are uncorrelated with each other. This is likely to be approximately true for real economies if our $N$ assets include not only traditional investment assets such as stocks and bonds, but also, e.g., the goods and services that make up the Consumer Price Index or the GDP. 

The second requirement is that an investment in a risk-free portfolio must not yield an economic gain. This means that the portfolio value must, in some sense, be constant. The naive notion of constancy is  to set the total time-derivative of the portfolio value to zero, but this fails to be gauge invariant. The simplest gauge invariant expression is

\begin{equation}
\label{constant}
\mathcal{D}\mathbf{P}\cdot\mathbf{K} = \left(\frac{d}{dt}+ A(t)\right) \mathbf{P}\cdot\mathbf{K}=0.
\end{equation}
Here $A$ is a time-dependent background gauge field that transforms as $A\to A^\prime=A-\dot{\phi}$ under the gauge symmetry, Eq.~(\ref{scaletransformation}).  It is straightforward to check gauge invariance by writing the equation in primed coordinates, substituting in the expressions in terms of the original coordinates, and verifying that all terms involving the gauge parameter, $\phi(t)$, cancel.  

Introducing the gauge field, $A$, solves the problem of gauge invariance, and it introduces a new degree of freedom that  allows a non-zero portfolio return as long as $A$ is chosen to satisfy Eq.~(\ref{constant}). We discuss the interpretation of the background gauge field in the next section.

The last requirement  is that risk-free portfolios  must be self-financing so that they can be used to translate asset flows in time. This is a refinement of the second requirement, as can be seen by writing   Eq.~(\ref{constant}) as
\begin{equation}
\label{chainrule}
\mathcal{D}\mathbf{P}\cdot\mathbf{K} = \left(\mathcal{D_P}\mathbf{P}\right)\cdot\mathbf{K} + \mathbf{P}\cdot \left( \mathcal{D_K}\mathbf{K}\right).
\end{equation}
The first term captures the value change due to price changes and the second term reflects in and out flows. Self-financing risk-free portfolios satisfy 
\begin{equation}
\label{costneutral}
 \left(\mathcal{D_P}\mathbf{P}\right)\cdot\mathbf{K}=0  \qquad\text{and}\qquad  \mathbf{P}\cdot \left( \mathcal{D_K}\mathbf{K}\right)=0,
\end{equation}
where the first equation is the requirement of constancy for self-financing portfolios. The second equation  requires that $\mathcal{K}$ represents a self-financing portfolio, because it requires all portfolio changes to be cost neutral.

If we require the two expressions to vanish separately, each  must be gauge invariant under the scale transformation, Eq.~(\ref{scaletransformation}), and  gauge transformations that redefine the unit of trade, Eq.~(\ref{qtransformation}). To ensure invariance under the latter, we introduce a new  $GL(M)\times GL(N)$  background gauge field\footnote{There is a larger symmetry group that allows redefinitions that mix options and underlying assets, but  the diagonal subgroup is sufficient for our purposes and avoids some subtleties.}

\begin{equation}
\mathcal{B} = \left( \begin{array}{cc} B_M & 0 \\ 0 & B_N\end{array} \right) 
\end{equation}
that transforms as
\begin{equation}
\label{btransformation}
{\mathcal{B}^\prime}^\gamma_\delta = b^\gamma_\kappa\mathcal{B}^\kappa_\rho{b^{-1}}^\rho_\delta - \dot{b}^\gamma_\kappa{b^{-1}}^\kappa_\delta.
\end{equation}
The two expressions in  Eq.~(\ref{costneutral}) are separately gauge invariant if we define the differential operators as 
\begin{align}
\mathcal{D_P}_\alpha^\beta &= \left(\frac{d}{dt}+A\right) \delta_\alpha^\beta+ \mathcal{B}_\alpha^\beta \\
\mathcal{D_K}_\alpha^\beta &= \frac{d}{dt} \delta_\alpha^\beta - \mathcal{B}_\alpha^\beta.
\end{align}
These two operators define what we mean by 'self-financing' and 'constant' in this paper. Combined with Eq.~(\ref{riskfree}) they define risk free portfolios.

Since all risk-free portfolios must satisfy Eq.~(\ref{costneutral}) with the same $A(t)$ and $\mathcal{B}(t)$, all must yield the same nominal return. For hedged portfolios the standard arbitrage argument that the risk free rate is unique applies. For infinitely diversified portfolios the arbitrage argument does not work because no finite transaction could have an effect on prices. In the next section we discuss the properties of asymptotically risk-free portfolios and show that they nevertheless all produce the same return.

\section{Asymptotically Risk-Free Portfolios}
\label{rfr}

Asymptotically risk-free portfolios are by construction insensitive to the specifics of each component asset in the $N\to\infty$ limit. Therefore their nominal return measures  global price inflation in the gauge implied by market data. Because this inflation rate applies to every asset, including cash, it has no economic impact even if the nominal return of these portfolios is non-zero. 

To see this formally, we evaluate Eq.~(\ref{costneutral}) on an infinitely diversified  portfolio
\begin{equation}
\label{qsevolution}
\left(\mathcal{D_P}\mathbf{P}\right)\cdot\mathbf{K}= \dot{\mathbf{s}}\cdot\mathbf{q} + A(t)  \mathbf{s}\cdot\mathbf{q} +  \mathbf{s}\cdot B_N\cdot\mathbf{q}=0
\end{equation}
and
\begin{equation}
\label{bfield}
 \mathbf{P}\cdot \left( \mathcal{D_K}\mathbf{K}\right)=\mathbf{s}\cdot\mathbf{\dot{q}}-\mathbf{s}\cdot B_N\cdot\mathbf{q}=0.
\end{equation}
The background gauge fields, $A$ and $B_N$ are unconstrained by symmetry arguments, but the evolution of $\mathbf{q}$ and $\mathbf{s}$ dictate their values. Solving Eq.~(\ref{bfield}) for $B_N$ we find 
\begin{equation}
(B_N)^i_j = \dot{q}^i (q^{-1})_j.
\end{equation}
This background field ensures that redefinitions of the unit of trade have no economic impact.
 Eq.~(\ref{qsevolution})  yields
\begin{equation}
\label{seta}
A(t)=- \frac{ \dot{\mathbf{s}}\cdot\mathbf{q} +\mathbf{s}\cdot B_N\cdot\mathbf{q}}{ \mathbf{s}\cdot\mathbf{q}} = - \frac{d}{dt}\ln(\mathbf{s}\cdot\mathbf{q}).
\end{equation}

Our definition of risk-free portfolios requires the background gauge field to be equal to the negative return of the risk-free portfolio. This ensures that the non-zero nominal return of the risk-free portfolio produces no economic benefit. We refer to $A(t)$ as the market gauge because it identifies the time-dependent global rescaling of market prices.

One way to understand the effect of $A(t)$ is to transform to the gauge where  $A^\prime=0$ by setting 
\begin{equation}
\phi(t) = \int^t d\tau A(\tau).
\end{equation}
In this gauge the background gauge field vanishes, and $\mathbf{K}\cdot\dot{\mathbf{P}}^\prime = 0$, indicating that the portfolio is constant. In the original gauge we have the illusion of a non-zero investment return if we ignore the effect of the background gauge field. 

To understand the effect of the background gauge field more generally, consider an arbitrary portfolio, $\pi$. Its time evolution is governed by the operator in  Eq.~(\ref{constant}), but since it is not necessarily risk-free there is no reason for $\pi$ to be constant. To calculate the return, $\mu$, of $\pi$ we write
\begin{equation}
\label{realreturn}
\mathcal{D}\pi = \left(\frac{d}{dt}+ A(t)\right)\pi = \mu\pi.
\end{equation}

The background gauge field subtracts the risk free rate (or equivalently the spurious return from global price inflation) from the nominal return of $\pi$ so that $\mu$ is the excess return over the risk free rate. Risk-free portfolios have zero excess return by construction to ensure that they do not yield an economic benefit. The real return, $\mu$, is gauge invariant and therefore a meaningful observable, while the raw portfolio return, $\dot{\pi}/\pi$, is not.  Raw returns like GDP numbers, consumer inflation, wage growth, etc.  need to be replaced with real returns according to Eq.~(\ref{realreturn}) to be meaningful.

The discussion so far uses the language of gauge symmetries and differential operators. Real prices have a random component and are usually modeled as stochastic variables. In order to connect our results to reality we promote the prices, $s_i$, to stochastic variables. 

Following \cite{gremm2015}, we assume that the parameters are functions of observables that characterize the market environment for a given asset. This leads to a process of the form 
\begin{equation}
\label{underlyingprocess}
ds_i = \mu_i(\xi(t))s_i dt + \sigma_i(\xi(t))s_i dZ_i,
\end{equation} 
where $\xi(t)$ represents one or more factors characterizing the market environment, and $dZ_i$ is random noise with zero mean, unit standard deviation, but not necessarily a normal distribution. The functions $\mu_i$ and $\sigma_i$ can be extracted from historical data given concurrent time series of both $s_i$ and $\xi$. More complicated processes may be required for certain assets, but this process should be general enough to capture the behavior of most assets.

By assumption all relevant environment variables are included in $\xi(t)$. They represent the in principle forecastable part of asset behavior. The remaining stochastic component must therefore be the unforecastable asset-specific noise. Any two unrelated noise processes should be uncorrelated: $\langle dZ_i, dZ_i \rangle = \delta_{ij}$.  The correlation in observed returns of similar assets arises because $\mu_i(\xi)$ for those assets  depends on $\xi$ in  similar ways. 

If the external parameters, $\xi(t)$, change slowly enough that they are approximately constant over the course of a day, the Central Limit Theorem states that daily returns are approximately normally distributed because they are the sum of returns on shorter timescales from approximately the same distribution. The process, Eq.~(\ref{underlyingprocess}) for daily data, becomes a standard log-normal random walk with parameters that depend on $\xi(t)$ and uncorrelated stochastic processes. The non-normality in observed time series of daily returns arises out of the time evolution of the environment variables, $\xi(t)$.  

This is not a simplifying assumption. It is a consequence of the parametrization adopted here that factorizes asset behavior into an (approximately) log-normal random walk with parameters that depend on $\xi(t)$ and a time series or forecast for $\xi(t)$.  See \cite{gremm2015} for a more detailed explanation, empirical tests, and concrete examples. 

Consider a portfolio $\Pi = \mathbf{q}\cdot\mathbf{s}$ with $q^i=\mathcal{O}(1/N)$ that becomes risk free as $N\to\infty$. The quantities are in general functions of the prices. We parametrize the quantities   in  terms of portfolio weights by setting $q^i = w^i/s_i$, where $w^i=\mathcal{O}(1/N)$ are a function of time only.  With these choices  the portfolio evolves as

\begin{equation}
\label{rfrtimeevolution}
\frac{d\Pi}{\Pi}= \frac{\mathbf{q}\cdot\mathbf{ds}}{\mathbf{q}\cdot\mathbf{s}}= \sum_i \left(w^i\mu_idt + w^i\sigma_idZ_i\right)=\hat{\mu}dt+\hat{\sigma}dZ
\end{equation}
where
\begin{equation}
\hat{\sigma}^2 = \sum_i w^{i2}\sigma_i^2, \qquad dZ = \frac{1}{\hat{\sigma}} \sum_i w^i\sigma_i dZ_i.
\end{equation}

Since $\hat{\sigma}=\mathcal{O}(1/\sqrt{N})$, the stochastic term vanishes as $N\to\infty$.  Integrating the resulting differential equation yields the risk free portfolio for all times. For finite $N$ the approximately risk free portfolio follows a log-normal random walk.

In the $N\to\infty$ limit any infinite subset of the $N$ assets furnishes a risk free rate. The deterministic term in Eq.~(\ref{rfrtimeevolution}) is a weighted average of the $\mu_i$. As shown in \cite{etemadi}, all of these weighted averages converge to the unweighted average of the asset returns provided the weights are positive and satisfy certain regularity conditions.  Consequently, all infinitely diversified portfolios with positive weights produce the same return.

Leverage is not available in risk-free portfolios. For example, if we own \$1 of a risk-free portfolio and borrow another \$1 to buy more of the risk-free portfolio, the resulting position is not risk free because we have a liability of \$1 against our \$2 of risk-free portfolio. The \$1 liability is a short position in a risky asset, the US Dollar, that does not scale as $1/N$. 

We could avoid this problem by borrowing a risk-free portfolio worth \$1 and exchanging it for more of the risk-free portfolio we already hold. Howver, this makes no economic sense. All risk-free portfolios yield the same and the lender would charge us at least the risk-free rate to borrow a risk-free portfolio, negating any benefit of leverage.

This does not preclude risk-free portfolios from contain shorts. In the US a \$2 long position can support \$1 short position without being on margin. As long as all weights scale as $1/N$ this is a  risk-free portfolio. However, the short portfolio is an asymptotically risk-free portfolio too. Its return is the negative of the asymptotically risk-free long portfolio, which makes any short position economically unattractive. We conclude that all asymptotically risk-free portfolios only contain  long position without leverage. This ensures that the convergence theorem in  \cite{etemadi} applies.

\section{Changes of Numeraire and  Gauge Transformations}
\label{numerairetransformation}

In the literature the freedom to choose a numeraire is often equated with the freedom to rescale prices according to Eq.~(\ref{scaletransformation}). These two symmetries are related, but they are not identical as can be seen by working  out how the price processes transform under them. In oder to explore these differences, we rewrite Eq.~(\ref{scaletransformation}) as 
\begin{equation}
s^\prime_i = e^\phi s_i=e^{\int^t d\phi}s_i= Ys_i
\end{equation}
and promote the gauge parameter to a stochastic process, $d\phi = \phi_\mu dt + \phi_\sigma dZ_\phi$, where $\phi_\mu$ and $\phi_\sigma$ are the drift and volatility respectively, and $dZ_\phi$ is normally distributed for simplicity. This makes $Y$ a log-normal random variable.  The price process in the primed coordinates reads
\begin{equation}
\label{numeraire}
ds^\prime_i =(\mu_i+\phi_\mu+\rho\sigma_i\phi_\sigma) s^\prime_i dt + \sqrt{\sigma_i^2 + \phi_\sigma^2 + 2\rho \sigma_i\phi_\sigma} s^\prime_i dZ_i,
\end{equation}
where $\rho$ is the correlation between $dZ_i$ and $dZ_\phi$.
Both the deterministic and the stochastic component of the price process changes under this transformation. 
To select $s_i^\prime$ as numeraire, we set $\phi_\mu = - \mu_i-\rho\sigma_i\phi_\sigma$, $\phi_\sigma = \sigma_i$, and $dZ_\phi = -dZ_i$, which ensures $\rho=-1$ and makes $s_i^\prime$ a constant. 

The ratio of two prices is invariant under this transformation. This suggests that this may be a symmetry of economies, in which case  the time evolution of a portfolio, $\Pi$, should be invariant  under numeraire transformations

\begin{equation}
Y\left( \frac{d}{dt} + A\right) \Pi = \left( \frac{d}{dt} + A^\prime\right) \Pi^\prime =  \left( \frac{d}{dt} + A- \frac{1}{Y}\frac{dY}{dt}\right) Y\Pi.
\end{equation}
However, 
\begin{equation}
\label{crossterms}
\left( \frac{d}{dt} + A- \frac{1}{Y}\frac{dY}{dt}\right) Y\Pi = Y\left( \frac{d}{dt} + A\right) \Pi + [Y,\Pi],
\end{equation}
where the last term represents the non-zero cross terms. The cross terms vanish if $\Pi$ has no stochastic term\footnote{They also vanish if the stochastic terms in $Y$ and $\Pi$ are uncorrelated but this does not hold for arbitrary $Y$.},  which is the case when $\Pi$ is insensitive to price changes. For example, if $\Pi  = V(s) + qs$ the portfolio is invariant under numeraire transformations  if $q=-\partial V/\partial s$, but not otherwise. 

Numeraire transformations affect the option price and the boundary condition. For example, the exercise  price, $E$, of an option becomes stochastic after a numeraire transformation. Using the standard homogeneity argument, the option  price can be written as an expansion in all quantities that transform under the symmetry
\begin{equation}
V(s,E) = c_s  s + c_E E,
\end{equation}
where $c_s$ and $c_E$ are functions of of  invariant ratios of the arguments. This parametrization  explicitly tracks the effect of numeraire changes on the underlying price and the boundary condition. 

To recover the non-stochastic gauge transformation from  Eq.~(\ref{numeraire})  we set $\phi_\mu = \dot{\phi}$ and $\phi_\sigma=0$. This turns off the stochastic component of $Y$, which causes the cross terms in Eq.~(\ref{crossterms}) to vanish even if $\Pi$ has a stochastic contribution. This  global symmetry applies to risky and risk-free portfolios alike. It does not change the stochastic properties of the price processes. A non-stochastic boundary condition remains non-stochastic under transformations. 

The invariance of $\sigma$  is  consistent with our earlier findings on portfolio returns. The volatility of a return stream is defined as 
\begin{equation}
\sigma^2 = \mathbf{E}\left( \left( \frac{\dot{\pi}}{\pi} -\mathbf{E}\left( \frac{\dot{\pi}}{\pi} \right) \right)^2 \right).
\end{equation}
Both the portfolios return and the expected value of the portfolio return transform as Eq.~(\ref{pitransform}) under gauge changes so the $\dot{\phi}(t)$ terms cancel. 

We can use the gauge symmetry to make an asymptotically risk-free portfolio the numeraire because it has no stochastic contributions. This is the natural gauge choice because sets the nominal return of risk-free portfolios to zero.

\section{Option Pricing}
\label{options}

Our definition of risk-free portfolios is a generalization of the arbitrage arguments that lead to the Black Scholes equation  \cite{black1973} combined with the requirement of gauge invariance. It is useful to see if we can recover the Black Scholes equation when applying our constraints to  a universe that contains $N\to\infty$ assets and one option, $V$, on the first  asset. Related ideas have been discussed in \cite{hoogland2001} and \cite{zhou2010}.

The option  singles out $s_1$ and breaks the unit of trade gauge invariance down to $R_+ \times GL(N-1)$. Per  Eq.~\ref{riskfree}, a price-insensitive portfolio must satisfy  $q^i=\mathcal{O}(1/N)$,  $i=2,\ldots,N$ and  the familiar delta-hedging condition,

\begin{equation}
\frac{q^1}{Q} =- \frac{\partial V}{\partial s_1}.
\end{equation}

The dynamics of $s_1$ are independent of those of $s_i$, $i=2,\ldots,N$. The entire portfolio is risk-free only if the two component portfolios are risk-free. Evaluating Eq.~(\ref{costneutral}) for both component portfolios and using Ito's lemma we find
%
\begin{equation}
\label{split}
\frac{\partial V}{\partial t}Q + \frac{1}{2}\sigma_1^2s_1^2\frac{\partial^2 V}{\partial s_1^2}Q - s_1 \frac{\partial V}{\partial s_1}QA +VQ\left(A+B\right)=0
\end{equation}
and
\begin{equation}
A = -\frac{ \dot{s_i} q^i + s_i (B_{N-1})^i_jq^j}{s_iq^i}.
\end{equation}

The second equation identifies $A$ with the negative risk free rate as before, except that the first asset no longer plays a role in setting $A(t)$. 

Eq.~(\ref{split}) is the Black Scholes equation with a new interpretation of the risk-free rate parameter as a background gauge field that corrects the time evolution of portfolios by removing the  effect of the global price rescaling. Instead of relying on arbitrage arguments, this derivation uses global symmetries and our definition of risk-free portfolios.

In the $N\to\infty$ limit, a portfolio given by  $q^i=\mathcal{O}(1/N)$,  $i=2,\ldots,N$ is risk-free. Consequently $A(t)$ is a deterministic function that we can set to zero via a gauge transformation. As we discussed in the previous section, gague transformations do not change the stochastic properties of the price processes. In particular, a non-stochastic boundary condition remains non-stochastic after a gauge transformation. 

We can impose boundary conditions in any gauge, but in order to compare option prices, it is necessary to impose boundary conditions in the same gauge for all options under consideration. For example, to meaningfully compare prices of options with different exercise prices, those prices need to be specified in the same gauge.  Here we adopt the arbitrary convention that boundary conditions are specified in the natural $A^\prime = 0$ gauge where the nominal and real return of a risk-free portfolio vanishes.

Real economies contain finite but large number of assets. These universes do not admit infinitely diversified risk-free portfolios. The only risk-free solutions are hedged portfolios. In order to accommodate this more realistic scenario we need to modify our requirements to allow approximately risk-free portfolios. The simplest choice is to relax Eq.~(\ref{riskfree}) so that it allows sensitivity to price changes of order $1/N$ while keeping the requirements that portfolios must be self-financing and bestow no economic benefit in place. 

With these assumptions, the option pricing equation for finite $N$ is given by Eq.~(\ref{split}) as before. However,  the background gauge field is now a stochastic variable given by the negative portfolio return from Eq.~(\ref{rfrtimeevolution}), $dA = -\hat{\mu}dt - \hat{\sigma}dZ$ with $\hat{\sigma}= \mathcal{O}(1/\sqrt{N})$. In order to impose boundary conditions in the $A^\prime=0$ gauge, we now need to make a numeraire transformation that moves the randomness from the stochastic part of $A(t)$ into the asset prices
\begin{equation}
s_i^\prime = e^\phi s_i = e^{\int^tdA} s_i.
\end{equation}
Using the stochastic process for $dA$ and the stock price, Eq.~(\ref{underlyingprocess}), the stochastic process for $s^\prime_i$ takes the form
\begin{equation}
ds_i^\prime = s_i^\prime\left( \mu_i-\hat{\mu} + \sigma_i dZ_i - \hat{\sigma} dZ\right),
\end{equation}
where we have neglected correlation terms that are suppressed by an extra factor of $1/\sqrt{N}$. Since the difference of normals is another normal, the price process is still log-normal with mean $ \mu_i-\hat{\mu} $ and standard deviation $\Sigma^2 = \sigma_i^2+\hat{\sigma}^2$.

The pricing equation in the primed variables reads
\begin{equation}
\label{bsprime}
\frac{\partial V^\prime}{\partial t}Q  + \frac{1}{2}\Sigma^2s_1^{\prime2}\frac{\partial^2 V^\prime}{\partial s_1^{\prime2}}Q - s_1^\prime \frac{\partial V^\prime}{\partial s_1^\prime}QA^\prime +V^\prime Q\left(A^\prime+B\right)=0
\end{equation}
with $A^\prime  = 0$. 
This is the same pricing equation as Eq.~(\ref{split}). The only effect of using an approximately risk free portfolio  is an $\mathcal{O}(1/\sqrt{N})$ bump in volatility from $\sigma_1$ to $\Sigma$. 

Solving Eq.~(\ref{bsprime}) with the same non-stochastic boundary condition as for Eq.~(\ref{split}) allows us to isolate the effect of replacing a risk-free rate with an approximately risk-free rate. If we imposed a non-stochastic boundary condition in a different gauge, this boundary condition would become stochastic in the $A^\prime=0$ gauge. The resulting solution would not be directly comparable to the a solution of  Eq.~(\ref{split}) because the boundary conditions do not match.

To further illustrate these concepts we briefly discuss the simplest option pricing equation with a stochastic `risk-free' rate,  \cite{merton1973}. In this model an option on a single asset is a function of the asset price and the price or a stochastic default-free bond. Homogeneity arguments provide a pricing equation, which can be solved with suitable boundary conditions.

In order to compare this model to our result we recreate it in the language of this paper. We assume that our economy has a finite number of assets, so that there is no risk-free rate.
Define this approximately risk-free portfolio as $H =\delta^{-1} \sum_{i=2}^N q^is_i$, where $\delta$ is an arbitrary scale $\mathcal{O}(1)$. This portfolio takes the place of the default-free bond in Merton's model. The option price takes the form $V(s_1,H,t)$. Using Eq.~(\ref{riskfree}), we find that this portfolio is has no sensitivity to price fluctuations if

\begin{equation}
\delta = - \frac{\partial V}{\partial H}.
\end{equation}
With this choice of $\delta$ we find
\begin{equation}
\label{merton}
\left(\mathcal{D_P}\mathbf{P}\right)\cdot\mathbf{K} = \left(\frac{\partial V}{\partial t} + \frac{1}{2}\sigma_1^2s_1^2\frac{\partial^2 V}{\partial s_1^2} + \frac{1}{2}\hat{\sigma}^2H^2\frac{\partial^2 V}{\partial H^2}+ \left(V-s_1 \frac{\partial V}{\partial s_1} - H \frac{\partial V}{\partial H} \right) \left(A+B\right)\right)Q=0.
\end{equation}
This is the gauge invariant extension of the pricing equation given in  \cite{merton1973}. 

The terms involving $A$ and $B$ are absent in  \cite{merton1973}.  We cannot simply set $A=B=0$. $A$ is the market gauge extracted from the observed return of the approximately risk-free portfolio, $H$. To set the terms involving background fields to zero we need to make a numeraire transformation, and possibly a trade unit transformation,  so that $A^\prime = A - \dot{\phi} = 0$ and $B^\prime=0$. The transformation that sets $A^\prime=0$ also sets the nominal and real return of $H$ to zero. This eliminates the terms involving gauge fields and reducing $H$ to a constant. The remaining dynamic terms reproduce Eq.~(\ref{bsprime}).  However, \cite{merton1973} imposes non-stochastic boundary conditions in the original gauge where $H$ is stochastic, while we impose them in the natural gauge. Even leaving aside the terms required for gauge invariance, the the option prices in  \cite{merton1973} are never comparable to ours because the boundary conditions are imposed in different gauges. 

The examples in this section illustrates the power of global symmetries. With minimal effort we were able to confirm that the Black Scholes equation respects them and that Merton's equation does not. It is rather remarkable that the Black Scholes equation is gauge invariant even though this was not a requirement built into the derivation. The Merton example shows that this is far from assured, even though his derivation is as plausible and logical as that of the Black Scholes equation. A possible explanation is that the Black Scholes world is so simple that there is only one way to price options. There are countless ways to incorporate stochastic 'risk-free' rates into the Black Scholes framework as the extensive literature on the subject shows. 

\section{Discounting Cash Flows}
\label{discount}

A seemingly simple application of risk free rates is discounting cash flows. The textbook method to value a  cash flow at time $T$ simply discounts the cash flow to the present using a possibly time-dependent risk free rate
\begin{equation}
\label{dcf}
c(0) =d(0,T) c(T)= e^{-\int_{0}^{T} dt r(t)} c(T).
\end{equation}
Here  $c= n\cdot s$ is a portfolio consisting of $n$ dollars with price $s=1$. 

There are several reasons why this method of discounting cash flows is problematic. It implies a risk-free profit if the risk-free rate is non-zero, which cannot happen if the economy is in equilibrium. In addition, the discount factor, $d(t,T)$, transforms as 
\begin{equation} 
d(0,T) \to  e^{\phi(0)-\phi(T)} d(0,T)
\end{equation}
under price rescaling. It economically meaningless because it is a gauge dependent. Finally, Eq.~(\ref{dcf})  ignores changes in the dollar's purchasing power, which can have a much larger effect on the real value of future dollars than the discount factor. 

In our framework we translate a dollar today forward by exchanging it for a risk-free portfolio, holding it until time $T$, and exchanging it for dollars at the then current exchange rate. Specifically, at time $t=0$ we buy $n_0 s(0)$ worth of a risk-free portfolio. Here $n_0$ is the number of dollars we invest and $s(t)$ is the price of a dollar. Since the risk-free portfolio is constant by construction, it is still worth  $n\cdot s(0)$ at time $T$. However, the  price or buying power of a dollar has changed. Consequently, when we do the exchange at time $T$ we receive a different number of dollars, 
\begin{equation}
n_0 s(0)=n_T s(T).
\end{equation}

The dollar, like all other assets, is stochastic. We do not know $s(T)$, but we can calculate the expected value. A portfolio, $\pi= ns(t)$, consisting of $n$ dollars with stochastic price $s$ evolves according to Eq.~(\ref{realreturn}). This stochastic differential equation  can be integrated analytically to yield

\begin{equation}
  n  \langle s(T)\rangle=  n s(0) \langle e^{\int_0^T dt\left( \dot{s}/s+A(t)\right)} \rangle = n s(0)e^{\int_0^T dt \left(\mu(t) -\frac{1}{2}\sigma^2(t) + A(t)\right)}.
\end{equation}
Reversing the direction so we discount from $T$ to the present as in  Eq.~(\ref{dcf}), we find 
\begin{equation}
\label{newdcf}
n_0 = \frac{\langle s(T)\rangle}{s(0)} n_T =  e^{\int_0^T dt \left(\mu(t) -\frac{1}{2}\sigma^2(t) + A(t)\right)}n_T.
\end{equation}
This discount factor is gauge invariant, because in involves a difference between a return and and a gauge field. It does not imply a risk-free profit because the dollar is risky. Eq.~(\ref{newdcf}) translates any asset, not just the dollar, in time. The discount rates are specific to the asset because the buying power of each asset evolves differently. Formally  Eq.~(\ref{newdcf}) reduces to Eq.~(\ref{dcf}) if we set $\mu=0$ and $\sigma=0$, but there is no such asset in our universe.

To illustrate these concepts we construct a crude approximately risk free rate out of 11 equally weighted tradable indices calculated in US dollars. Our dataset covers the decade from July 2005 to June 2015. We arbitrarily set the price per share of the approximately risk free portfolio on June 31st, 2005 to \$1 so that the dollar is worth one unit of the risk free portfolio. We also arbitrarily  scale the price per share of all other instruments to that they are \$1 on June 31th, 2005.

The non-zero return of our 11 index portfolio determines $A(t)$. It is convenient to recast the data in the gauge where the backround field vanishes and the approximately risk-free portfolio is constant. This is easily done by expressing the data in units of the approximately risk-free portfolio. In these units the approximately risk-free portfolio is constant, but the dollar and the 11 indices are stochastic. 

To effect this conversion we simply divide the dollar price per share of each asset by the dollar price per share of the risk free portfolio. For the dollar we divide \$1 by the price per share of the risk free portfolio. Given our pricing scheme, all 12 instruments were worth one unit of the approximately risk-free portfolio  on June 31st 2005.  The value of all 12 instruments in units of the risk free rate on June 31st, 2015 is listed below.

\[ \begin{tabular}{|l|r|}
\hline
\multicolumn{2}{|c|}{\bf Final Asset Values} \\
\hline
MSCI Emerging Market Stock		& 1.252 \\
S\&P 500 					& 1.226 \\
MSCI EAFE					& 0.945 \\
\hline
Barclays High Yield				& 1.220 \\
JPMorgan Emerging Markets Bond	& 1.183 \\
iBoxx Liquid Investment Grade		& 0.972 \\
Barclays Broad Bond			& 0.886 \\
Barclays Inflation Linked Bond		& 0.864 \\ 
\hline
DJ US Real Estate				& 1.018 \\ 
DJ Global ex-US Select Real Estate	& 0.987 \\
\hline
US Dollar					& 0.574 \\
\hline
S\&P GSCI Commodities			& 0.314 \\
\hline
\end{tabular} \]

\begin{figure}[h]
\center
\includegraphics[width=100mm]{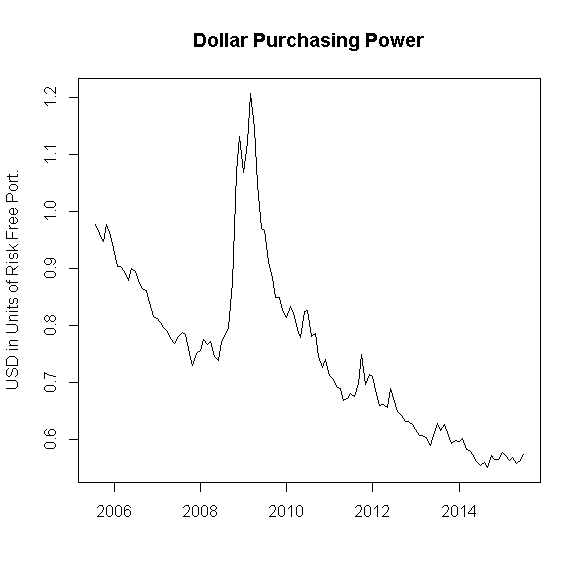}
\caption{$s(t)$ for the Dollar in the gauge where $A^\prime=0$}
\label{usdplot}
\end{figure}

Since our approximately risk free portfolio has a positive return in dollar terms, the dollar in units of the risk free rate has a negative return, reflecting its generally declining buying power.  However,  Fig.~\ref{usdplot} shows that during the 2008 crisis the purchasing power rose sharply before resuming its decline. 

We can calculate the discount factors for all assets using Eq.~(\ref{newdcf}) and the final prices in the table above. For the dollar we find that $n_T=1$ dollar on June 31st, 2015 is equivalent to $n_0 = 0.574$ dollars on June 31st, 2005. If we discount at the 10-year government yield of 4.06\% on June 31st, 2005, we find that $n_0=0.666$ instead. 

Conceptually the calculations are very different, but at least for our small data sample, the results are comparable. It remains to be see if this holds true with more carefully constructed approximately risk-free portfolios and longer time periods. 

\section{Parameter Sensitivities}
\label{sensitivity}

The results presented so far assume that we know the value of the  parameters $\xi(t)$ for all times. This is a common assumption in the literature. For example, the textbook expression for discounting cash flows, Eq.~(\ref{dcf}), assumes that $r(t)$ is a known function. Similarly, the Black Scholes equation assumes that the parameters of the price process are known. 

These assumptions are almost never satisfied in reality. We do not know the future evolution of the risk-free rate with certainty. Nor do we know the value of the forward-looking volatility that determines the Black Scholes option price. 

A full discussion of these uncertainties is beyond the scope of this paper, but the general approach to incorporate parameter uncertainty consists of obtaining a probability distribution of the parameter values and taking expectation values of quantities that depend on these parameters. For example, an option price calculated in this manner would reflect not only the volatility of the underlying, but also the risk associated with uncertain parameters. 

In our framework the parameters of the stochastic processes are functionals of $\xi(t)$. As shown in \cite{gremm2015}, these functionals are stable and can be estimated from historical data. The main source of parameter uncertainty stems from the forecasting error associated with $\xi(t)$ for future dates. This drastically reduces the number of sources of uncertainty.  Instead of dealing with a few uncertain parameters, e.g.,  $\mu$ and $\sigma$, for each of $N\to\infty$ assets, we only need to deal with the uncertainty in a finite number of factors characterizing the economic environment. 

For $N$ asset portfolios it is often possible to reduce the exposure to parameter uncertainty. For example, the asset weights, $w_i=\mathcal{O}(1/N)$, for infinitely diversified portfolios are not uniquely fixed by the requirement that the portfolio is risk-free. We can chose them so that the expected return is insensitive to variations, $\xi(t)\to\xi(t)+\delta\xi(t)$. In most cases there should be a choice of weights that satisfies the risk-free condition and 
\begin{equation}
\delta\hat{\mu} = \sum_i w^i \frac{d\mu_i}{d\xi(t)} \delta\xi(t) = 0.
\end{equation}
These portfolios are insensitive to small forecasting errors and should yield more reliable forward-looking risk-free rates. 

Note that the weights of a delta-hedged portfolio of an option and the underlying asset are fixed up to an overall scale by the risk free condition. There is no remaining freedom that we can use to also make them insensitive to parameter uncertainty. Consequently realistic option prices should always reflect the risk associated with parameter uncertainty.

\section{Conclusions}

We proposed a definition of risk-free portfolios that respects global symmetries of economies in equilibrium. It can be expressed via three differential operators that send risk-free portfolios to zero.  Under this definition the default-free bonds are risky, but hedged options and  infinitely diversified portfolios are risk-free. The return of the latter is the risk-free rate, which measures an economically meaningless  global rescaling of all prices.  

A non-zero risk-free rate induces a background gauge field associated with the price rescaling symmetry that subtracts the nominal risk-free rate from every other return. This ensures that the real return of risk-free portfolios is zero and that all other returns are measured relative to the risk-free rate. For example, if we are interested in GDP growth or the performance of the S\&P 500, we need to calculate the difference between these  nominal returns and the risk-free rate to strip out the component of the nominal return that is due to global price rescaling. The resulting real return is a gauge invariant quantity that is economically meaningful.

In practice it is straightforward to compute real returns. A good approximation of the risk free rate is the return of an equally weighted portfolio of as many assets as possible. Infinitely many assets would yield a true risk-free rate, but we can construct a  close approximation in real economies because they contain a  large number of assets including stocks, bonds, consumer goods, commodities, services, etc. 

Our framework provides a new interpretation of the risk-free rate as a global price rescaling. It reproduces  the Black Scholes equation and provides extensions to approximately risk-free discount rates. We also discuss how to discount cash flows using our definition.

In this paper we applied symmetry arguments in the very limited context of risk-free portfolios to gain a better understanding of the risk-free rate and to provide a few examples of how symmetry arguments work in finance. However, symmetry arguments apply much more broadly. Empirical researchers and policy makers need them to determine which quantities are meaningful observables, model builders need them to ensure that their models respect global symmetries, and investors need them to determine the real return of assets. 

\section{Acknowledgment}

The author thanks the Walter Burke Institute at Caltech for hospitality, and Mark Wise, Anton Kapustin, and Didier Sornettee for insightful discussions and comments. 

\bibliographystyle{ieeetr}
\bibliography{../Distributions/finance}

\end{document}